# An Exploratory Philosophy of Statistical Data

Putting the Data in Data Ethics

Alexander Martin Mussgnug
05/19/2020

# 1 Introduction

In 2017, The Economist headlined, "The world's most valuable resource is no longer oil, but data" (The Economist, 2017). Data are not only increasing in value but also affect our lives in the most profound ways — from guiding our love life (Rajan, 2019) to dictating the value of each citizen in China's Social Credit System (Liang et al., 2018). The field of data ethics seeks to investigate the social implications and moral problems related to data and data technologies (Floridi & Taddeo, 2016). While this discourse is of critical importance, we argue that it is missing one fundamental point: If more and more efforts in business, government, science, and our daily lives are data-driven, we should pay more attention to what exactly we are driven by. To this end, we need more debate on what fundamental properties constitute data.

In its everyday use, the term "data" is of considerable variability, depending on the domain and situation. Therefore, we face the question of which concept of data might be most fruitful for the debate on data ethics. Data ethics has developed into a broad area of research that not only asks questions about data, its generation, storage, or dissemination but also investigates the uses of data, for instance, in the form of algorithms (Floridi & Taddeo, 2016; Mittelstadt et al., 2016). More fashionable terminology, such as artificial intelligence or data science, might obscure the fact that especially when looking at the applications of data, these practices often fall within the realm of statistics. Therefore, we argue that in these cases, data ethics can, to some degree, be understood as an ethics of statistics. At least to the same extent, we can identify the conception of data at play, as one of statistical data.

In the first section of the paper, we work from the fundamental properties necessary for statistical computation to a definition of statistical data. We define a statistical datum as the coming together of substantive and numerical properties and differentiate between qualitative and quantitative data.[1] Subsequently, we qualify our definition by arguing that for data to be practically useful, it needs to be commensurable in a manner that reveals meaningful differences that allow for the generation of relevant insights through statistical methodologies.

In the second section, we focus on what our conception of data can contribute to the discourse on data ethics and beyond. First, we hold that the need for useful data to be commensurable rules out an understanding of substantive properties as fundamentally unique or fundamentally equal. Instead, useful data must be defined at a level of generality between the two extremes. Second, we argue that practical concerns lead us to increasingly standardize how we operationalize a substantive property; in other words, how we formalize the relationship between the substantive and numerical properties of data. Thereby, we also standardize the interpretation of a property. With our increasing reliance on data and data technologies, these two characteristics of data affect our collective conception of reality. Statistical data's exclusion of the fundamentally unique and equal influences our perspective on the world, and the standardization of substantive properties can be viewed as profound ontological practice, entrenching ever more pervasive interpretations of phenomena in our everyday lives.

---

[1] For the remainder of this paper, we will generally refrain from repeating the qualifier "statistical," and instead refer to data.

# 2 Ontology of Data

## 2.1 Existing Definitions of Data

Before we define data within the field of statistics, we will outline existing conceptions of data in philosophy. These approaches to understanding data are more general and might prove more valuable for equally general matters or instances where data are not statistical. Presumably, the most fundamental approach to a definition of data can be found in Floridi (2011), where data are understood as the raw material of information. In his philosophy of information, "a datum is ultimately reducible to a lack of uniformity." A datum is "x being distinct from y," where x and y remain uninterpreted (Floridi, 2011, p. 85). If data are well-formatted (syntax), meaningful (semantics), and truthful, they become information (Floridi, 2011b, p. 104). One might question whether this definition aligns with the meaning of the term "data" in everyday language or within statistics. Instead, it might be considered an ontological argument about the nature of reality and information.

Within the philosophy of science, Leonelli (2016) distinguishes between representational and relational accounts of data. A representational account puts forth a context-independent underlying definition of data. One example of a representational interpretation is the definition proposed by the Royal Society (2012, p. 12) of data as "numbers, characters or images that designate an attribute of a phenomenon." It is closely related to notions of data as "raw" materials of research and a realist view on measurement. Leonelli (2015) herself advocates for a relational view on data, arguing that what is considered data is context-dependent. More precisely, whether something constitutes data is contingent on the use case as long as it is portable and potentially useful as evidence.

The primary purpose of this article is not to define data in general or within science as a whole, but to give a definition of data within statistics and to illustrate how this can help ground social and ethical issues specific to data of such kind. Nonetheless, we view our work in support of the thesis that what we regard as data depends on the context. By limiting ourselves to statistical data, we are effectively providing such a context in the form of the standards and methodologies that limit what we can use as evidence within statistics. Consequently, our definition might not be entirely applicable to other areas of science. While we believe that further research contrasting notions of data between disciplines could provide meaningful insights into the field's epistemology and scope, we limit our analysis to the field of statistics.

## 2.2 Defining data

We shall briefly outline some definitions of statistics itself before seeking one of statistical data. Savage (1977), for instance, defines statistics as the study of uncertainty. Others define statistics as "the practice of gathering data and making inferences based on them" (Bandyopadhyay & Forster, 2011, p. 1). Agresti and Franklin (2009, p. 4) also identify statistics as "the art and science of learning from data." As we lack a precise understanding of statistical data, such proposals leave open the substantive question of what exactly we are learning from. If we concede that those data-based definitions of statistics carry some truth, we can see how crucial an investigation into the nature of statistical data can be to better understand the epistemology of statistics, as well as its social and ethical implications.

By looking at various definitions of statistics, it becomes evident that we can approach statistics from different angles, as well as understand its domain more or less broadly. Statistics might include elements such as observation, information collection, preprocessing, computation, visualization, and interpretation (Freund & Miller, 1965). These different processes depend on different kinds and representations of information. The information gained at the point of observation, for instance, can be vastly different from the information before computation in both meaning and representation. If we want to define data as the informational material of statistical analysis, we face the question of where to locate statistical data.

For our purposes, we identify data as information that can serve as the input for statistical calculation. We have three main reasons for doing so. First, we want statistical data to be sufficiently distinct from information in general so that our definition is relevant for questions specific to statistics. While almost all information can be collected, processed, visualized, or interpreted, the information admissible for calculation is of a specific format. Second, while the gathering and processing of information can make up significant parts of a statistician's work, a statistician does not collect or process data as an end in itself. He or she does so to enable insights, even if these acts are temporally distant or carried out by different individuals. These insights are most often gained through visualization or computation. Lastly, our definition of data should give us relevant insights on social and ethical issues. We believe that inferences and ultimately actions, based on statistical computation, including machine learning and other data technologies, are of considerable social and moral significance. All these factors lead us to pinpoint data as the information that is admissible for statistical computation. Locating data at this place does not imply that only information immediately before computation is considered data, but that everything we consider to be data would be directly admissible for such calculation. The information found at other stages of a statistician's work might equally qualify as data, if admissible for calculation. In other cases, we might understand it as a predecessor of data. Readers objecting to our focus on statistical computation may qualify our definition as not about statistical data, but data for statistical computation.

To arrive at our definition of a datum, we must look at the most fundamental properties that enable statistical calculation. Calculation is numerical. It follows that data must have numerical properties to be admissible for calculation. However, these numerical properties are not merely abstract numbers. To allow for representation or insights into more than the nature and relationship of numbers, statistical data must have substantive properties. The numbers must represent a substance, whether this substance is known ex-ante or still unidentified. These substantive properties are what allows for data's "prospective usefulness as evidence" (Leonelli, 2015, p. 2) or "designate an attribute of a phenomenon" (Royal Society, 2012, p. 12). We conclude that for data to be admissible for statistical computation, it needs to have numerical and substantive properties. Consequently, we arrive at our definition of a statistical datum:

*A statistical datum is a numerical property representing a substantive property of a specific object.*

## 2.3 Types of Data

Qualitative Data

Recall that we identified data as the kind of information that one can immediately apply in statistical computation. There, it becomes the value of a variable. These variables can be qualitative (also: categorical) or quantitative. For this reason, we can also distinguish between qualitative and quantitative data. The categories of qualitative data are qualitative substantive properties that we can assign to objects. At times, we represent these categories by numbers that function as labels. These labels are, as Espeland and Stevens (2008, p. 407) note, "numbers that mark." Such labels can be arbitrary and do not need to be numerical at all. A letter or other symbol can equally function as a label. Therefore, they should not be confused with the numerical properties of data that enable statistical computation. To identify the numerical properties of qualitative data, let us imagine a simple dataset on two students' athletic activities. Joe swims and cheers, while Jessica plays soccer. The numerical property of the individual datum is the binary numerical answer to the question, "Does object x have property y? ". In our case, we could ask, "Does Joe play Basketball?". By its numerical property, qualitative data records the presence (1) or absence (0) of such a qualitative property.

**Table 1**

*Qualitative Data on High-School Athletes*

|         | *Basketball* | *Swimming* | *Cheerleading* | *Soccer* |
|---------|--------------|------------|----------------|----------|
| Joe     | 0            | 1          | 1              | 0        |
| Jessica | 0            | 0          | 0              | 1        |

**Table 2**

*Numerical Value and Substantive Property*

|              | *Plays Basketball (Substantive Property)* |
|--------------|-------------------------------------------|
| Joe (Object) | 0 (Numerical Property)                    |

Quantitative Data

While qualitative data only identifies the presence or absence of a property, the substantive properties of quantitative data allow for comparison by degree. It is not by having numerical properties that quantitative data gets its name, as both quantitative and qualitative data have numerical properties. Instead, the qualifier "quantitative" refers to the substantive property of the datum. With quantitative data, we can now compare the property by degree. For instance, we can say not only whether Jessica plays soccer or not (qualitative), but how many goals she shot in comparison to her teammates (quantitative). Moreover, the numerical properties of quantitative data can also include ranges or distributions of numbers.

## 2.4 Contouring Data

Our definition of a datum as a numerical property representing a substantive property of a specific object is very close to the one put forth by the Royal Society (2012, p. 12), which defined data as "numbers, characters or images that designate an attribute of a phenomenon." Together they paint an image of statistical data as a specific case of scientific data. However, it is also not in contradiction with the definition put forth by Leonelli (2015). When data were to be context-dependent, wouldn't it permit differentiating data in diverse contexts? Within a relational conception of data, our definition presents a context-specific subtype of data. Still, the question remains, how data are distinct from information.

It is evident that not every information immediately also constitutes data. This text, as an example of linguistic information, does not have explicit numerical properties. It is also correct that even if not explicitly represented, we can easily find numerical properties, such as the number of words. However, the question we are facing is whether all information can be translated into data, which would make data merely a specific representation (syntax) or whether only specific information can be translated into data, rendering data specific in meaning (semantics). To answer this question, we would have to analyze it in light of different conceptions of information. We acknowledge the need for further research into this issue. At the same time, we wish to move from rather theoretic to more practical considerations in order to illustrate how an explicit analysis of the properties of data can be of value for social and ethical debates.

## 2.5 The Need for Commensurability

Up until now, we based our analysis on the essential properties of data. What we described so far is the bare minimum, theoretically, enabling something to be used as a statistical datum. We paid no attention to the question of whether the data is meaningful or suitable. Now, we will turn our heads to the property that makes data practically useful. Statistical calculation is numerical. As Jevons (1874, p. 177) notes, a number "consists in abstracting the character of the difference from which plurality arises, retaining merely the fact." Consequently, by being about numbers, statistics is about differences. More precisely, it is the study of how differences and equalities of one kind, abstracted as numbers, relate to differences of another kind, usually differences in probability. When calculating, we extract the numerical properties from the data, retaining only the relationships of difference or equality that are the numbers and (quite literally) the units. Statistical computation is a fundamentally relational practice, not concerned with the substantive properties themselves but only with their numerical relationships. Therefore, we must understand the usefulness of data in equally relational terms. A datum is not useful in and of itself but relative to the other data available and the task at hand. A single datum is only useful for statistics if commensurable in a manner that reveals differences (relative to other data) that allow for the generation of relevant insights (relative to the use case) through statistical methodologies:

*Useful statistical data are numerical properties representing substantive properties of specific objects that are commensurable in a manner that reveals meaningful differences that allow for the generation of relevant insights through statistical methodologies.*

By now, we defined statistical datum, differentiated between qualitative and quantitative data, and identified that for data to be useful, it needs to be commensurable in a way that reveals meaningful differences. This need for data to be commensurable in order to be useful can help us ground at least two issues of social and ethical significance.

# 3 Ethics of Data

## 3.1 Data, the unique, and the equal

Let us start with an example illustrating how the level of generality with which we consider a substantive property influences the usefulness of data. People might have very distinct and complex mental abilities that we label intelligence. Goethe might have been able to capture his readers with his exceptional linguistic abilities, while Beethoven's unique musical genius overshadows his more questionable social intelligence. In the same fashion, all people have very individual combinations of highly specific abilities that we associate with intelligence. When seeking data for statistical computation, however, we would hardly identify each of these sets of abilities at a level of detail where it applies to only one person. It would also make little sense to define intelligence so that it applies to all of them in totally equal terms. Instead, we would seek categories that allow for relevant groupings, such as "musically talented" or "analytically gifted" (qualitative) or general properties that apply to all of them in varying degrees, such as the IQ score (quantitative).

As previously defined, useful statistical data entails defining the substantive properties in a manner that reveals meaningful differences, allowing for the generation of insights through statistical methodologies. But what exactly does it mean to define the substantive properties in such a fashion? First, our example illustrates that it is of little use to view the substantive properties at a level of specificity at which the property only applies to a single object, such as a holistic and very detailed view on an individual's intelligence. Remember that when calculating with the numerical properties of the data, we discount the substantive property itself, retaining merely the relationships of difference and equality. When we define a property as unique to a single object, all that one can statistically infer is that the object is distinct from all other objects. The value of this insight is, at best, limited. The fact that an object is distinct is self-evident, and no further substantive property defined in such a fashion adds any benefit to statistical computation. The same applies to properties so general and broad that they apply to all of them to the same degree. If all objects are equal with respect to that property, we have no difference that we could base our statistical computation upon. We can see that there is an ideal level of generality that is neither too detailed to the point where something is unique nor too general, where no differentiation is possible. Where we define a property on this spectrum between generality and specificity is subject to the application, epistemic access, and statistical methodology. However, what we can say about statistical data, in general, is that they are not useful when defined at the extremes of total uniqueness and total equality.

**Figure 1**

*Usefulness of data as a function of generality*

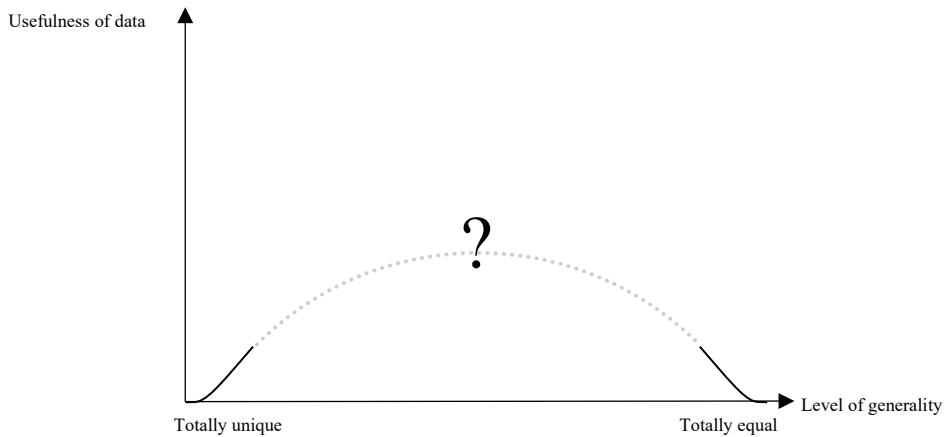

## 3.2 Data and ontological standardization

We illustrated how the need for statistical data to be commensurable rules out the two extreme understandings of phenomena as fundamentally unique or fundamentally equal. Between these two extremes lies the space where useful statistical data are possible. As mentioned, where precisely we define a property on this spectrum between generality and specificity is subject to the application and methodology. However, we will now illustrate that practical considerations arising from the need for commensurability, often lead us to standardize our understanding of a property.

Let us introduce another, more empirical example that will help us better understand this second consequence of data's need for commensurability. The Binet-Simons Intelligence Scale, later revised as the Stanford-Binet Scale, is the origin of the intelligence quotient (IQ) as we know it today and profoundly influenced our conception of intelligence. As Carson (2007, p. 159) notes, "out of a variety of different ways of conceptualizing intelligence in play at the beginning of the century, one dominant theme had emerged. Intelligence was understood as a differential, quantifiable, unilinear entity that determined an individual's or group's overall mental power." The IQ score is to this day quasi-synonymous with intelligence, standardizing our interpretation of it. Furthermore, it is an example of information in the format of statistical data. The IQ score is a number (numerical property) representing the intelligence (substantive property) of an individual or group (object), rendering the concept of intelligence commensurable in a manner that allows for meaningful statistical analysis.

The Binet-Simons Intelligence Scale also illustrates how we can interpret the generation of data as ontological practice. The IQ score seeks to measure intelligence as a substantive property. However, as noted by Carson (2007), intelligence is a concept that we can understand in many diverse ways. By defining a way of measuring it, it formalizes the relationship between the numerical and substantive properties of the datum. In other words, the IQ score formalizes how to put a number (numerical property) on intelligence (substantive property) and, thereby, defines what counts as intelligence. This interpretation, in turn, becomes the operational definition of the concept at the basis of statistical data and inferences made based on them. In this manner, we can understand the generation of data as ontological practice. It similarly applies to qualitative data in the form of the rules and procedures for assessing whether an object does or does not belong (numerical property) to a specific category (substantive property).

The critical issue we face here is not the fact that we interpret a substantive property in one way or the other. Instead, we are concerned with the fact that practical concerns for commensurability incentivize us to standardize such interpretation. There is much truth in the saying that one cannot compare apples to oranges. If we were to have a multitude of different metrics, and, therefore, interpretations of a property, we would face the problem that the data might not be commensurable. Through having one standard interpretation, we can most easily make use of the most data available. For this practical and economic reason, we often seek to standardize our measurement practice and, thus, our conception of phenomena.

## 3.3 Datafication and our conception of the world

As outlined, two consequences arise from data's need for commensurability. First, useful data must be defined at a level of generality that excludes the interpretation of phenomena as fundamentally unique or equal. Second, practical concerns for commensurability incentivize us to standardize our interpretation of substantive properties. What makes these facts of social and ethical relevance is the increasing proliferation of data and data technologies, also called datafication (Cukier & Mayer-Schoenberger, 2019). The need for discussion of these characteristics of statistical data in a social and ethical context arises from the ever-increasing proliferation of statistical data in our everyday life and the implicit influence it has on our understanding of reality. Statistical data are increasingly at the core of systems we inhabit, at the core of the decisions we make, at the core of our scientific knowledge, and, ultimately, at the core of our everyday experience of the world. With this growing importance of data in our lives, the characteristics of statistical data have the potential to alter our interpretation of ourselves and the world around us.

Useful statistical data omits the unique and equal. Our increasing reliance on statistical data might also lead us to lose sight of the unique and fundamentally equal in our social and individual conception of the world. Not because we explicitly choose to do so but because we increasingly inhabit systems and domains of knowledge that are data-driven. Therefore, we are inclined to understand the world around us in a manner that aligns with the interpretation of substantive properties essential to useful statistical data: an understanding of objects as defined by shared properties and their belonging to general categories. Moreover, it is crucial to keep in mind that when we define the substantive properties of data at a certain level of generality, information about all more specific differences is lost. That means that in all places where data dominates our understanding, the qualitative information that renders something unique is lost. This subtle change in our conception of the world brought about by datafication deserves our attention.

Arguably more tangible is the standardization of our understanding of substantive properties. A shared metric enables us to effectively commensurate data from different sources. Especially since we often use secondary data, concerns for commensurability with exiting data play an important role in statistical work and incentivize us to standardize our interpretations.

This standardization, however, comes at the expense of the plurality of conceptions and must be critically reviewed for its cultural implications and power structures involved. The growing reach of this homogenization of interpretations does not stop with data. With the proliferation of data technologies, the desire for standardization of interpretations in data will implicitly and explicitly standardize the conceptualization of phenomena in our daily experience. The decisiveness and social relevance of this dynamic can hardly be overstated. Again, we see how the need for commensurability plays a crucial role in the way data can shape our understanding of the world around us.

Commensurability and ensuing issues of generality and standardization are not necessarily novel, but tendencies existent in science in general. Nevertheless, they are hardly as definitive as with statistical data and as expansive as with the current datafication, profoundly transforming science, industry, government, and the general society. They are also more complex than we could analyze here. Not only do they interrelate with each other, but they are also neither strictly good nor bad. A perspective more focused on shared and distinct features might have integrative tendencies and allow for insights otherwise not possible. Nevertheless, we might question whether the influence this has on our social and individual reality is strictly positive. Standardization, on the one hand, also has undeniable benefits for the advancement and costs of knowledge. On the other hand, it disincentivizes the adaption of other conceptions of phenomena, restricts our understanding of them, and gradually establishes a definite interpretation in our daily experience through the expansion of data and data technologies.

There remains much work to be completed. What we have done so far is neither definite nor exhaustive. Instead, we sketched one possible conception of statistical data in a manner that can only be exploratory. More than anything, our goal is to illustrate that such an explicit treatment of the fundamental properties of data can be of value for debates about data ethics and beyond. We also hope to motivate further research into how conceptions of data differ amongst disciplines and from information in general, as well as how they relate to ethical and social concerns raised. Not only concerns about specific kinds of algorithms should entertain our interests. The ways in which a data-driven society changes our perception of and place in the world, though more subtle, are by no means less decisive. For this reason, we must make an effort to better understand the characteristics of data.

# 4 Conclusion

In this paper, we provided one possible definition of data and showed how we can put it to work. In the first section, we defined a datum as the coming-together of substantive and numerical properties and differentiated between qualitative and quantitative data. We qualified this definition by arguing that data is only useful for statistics if commensurable in a manner that reveals meaningful differences that allow for the generation of relevant insights through statistical methodologies. In the second section, we focused on what our conception of data can contribute to the discourse on data ethics and beyond. First, we held that the need for useful data to be commensurable rules out an understanding of substantive properties as fundamentally unique or fundamentally equal. Instead, useful data must be defined at a level of generality between the two extremes. Second, we argued that practical concerns lead us to increasingly standardize how we operationalize a substantive property; in other words, how we formalize the relationship between the substantive and numerical properties of data. Thereby, we also standardize the interpretation of a property. With our increasing reliance on data and data technologies, these two characteristics affect our conception of reality. Statistical data's exclusion of the fundamentally unique and equal influences our perspective on the world and the standardization of substantive properties can be viewed as profound ontological practice, entrenching ever more pervasive interpretations of phenomena in our everyday lives. However, more than anything, our goal is to demonstrate why we need an intensified debate about what exactly data is. If our collective and individual activities and reality become increasingly data-driven, we should talk about what it is we are driven by.